\documentclass{article}

\usepackage[authoryear]{natbib}
\usepackage{amsmath, amssymb}
\usepackage{graphicx}
\usepackage{booktabs}
\usepackage{hyperref}
\usepackage{geometry}
\title{Gaussian Field Representations for Turbulent Flow: Compression, Scale Separation, and Physical Fidelity}

\author{
Dhanush V. Shenoy \\
Faculty of Mechanical Engineering \\
Technion -- Israel Institute of Technology \\
\texttt{dhanushv@campus.technion.ac.il}
\and
\vspace{0.3cm}
Steven H. Frankel \\
Faculty of Mechanical Engineering \\
Technion -- Israel Institute of Technology
}
\date{}

\begin{document}

\maketitle

\begin{abstract}
Representing turbulent flow fields in a compact yet physically faithful form remains a central challenge in computational fluid dynamics. We propose a continuous parametric representation based on localized Gaussian primitives, in which the velocity field is modeled as a superposition of kernels with learnable positions, amplitudes, and characteristic scales. This formulation yields a compact, grid-independent encoding while enabling analytical evaluation of derived quantities such as vorticity and enstrophy.

The approach is assessed on three-dimensional Taylor--Green vortex fields spanning different stages of flow evolution, from smooth configurations to fully developed turbulence. We quantify the compression--accuracy trade-off using both primary flow variables and derivative-sensitive diagnostics. The baseline isotropic formulation achieves high velocity reconstruction accuracy at compression ratios exceeding $10^3$--$10^4$, but exhibits substantial degradation in enstrophy due to the loss of small-scale turbulent structures.

To address this limitation, we investigate several structure-aware extensions, including adaptive kernel placement, multi-resolution kernel distributions, and anisotropic Gaussian kernels. Among these, the anisotropic formulation provides the most consistent improvement, enabling better alignment with elongated vortical structures and improved recovery of intermediate- and high-wavenumber content. Adaptive and multi-resolution strategies yield only modest gains under comparable conditions. We also examine an alternative compact-support Beta basis, which can improve enstrophy recovery in some cases but introduces localized reconstruction artifacts and reduced smoothness.

Overall, the results show that the principal limitation of baseline Gaussian representations lies in their geometric expressiveness rather than in parameter count alone. The proposed framework provides a compact, interpretable, and continuous representation of turbulent flows, and establishes a foundation for structure-aware and physics-informed approaches to compact, continuous representations of turbulent flows.
\end{abstract}



\section{Introduction}

Representing turbulent flow fields in a compact yet physically faithful manner remains a central challenge in computational fluid dynamics. Turbulent flows contain a wide range of interacting spatial scales, intermittent coherent structures, and regions of strong vorticity~\cite{pope2000,frisch1995}. Faithfully capturing such behavior typically requires dense discretization and large data volumes, creating challenges for storage, transmission, and analysis. These challenges have motivated the development of reduced-order modeling and data-driven approaches for flow representation, including modal decompositions and machine-learning-based methods~\cite{lucia2004reduced,rowley2005model,brunton2020machine,thuerey2020deep,Momenifar2022}. Developing representations that are both compact and capable of preserving physically relevant flow features is therefore of substantial practical and scientific interest.

Classical reduced-order approaches provide a natural starting point for compressing high-dimensional flow data~\cite{benner2015survey}. Projection-based methods such as Proper Orthogonal Decomposition (POD) and Dynamic Mode Decomposition (DMD) represent the flow using global bases~\cite{berkooz1993proper,schmid2010dynamic}. While effective for capturing dominant large-scale dynamics, their global nature limits their ability to resolve localized, multiscale turbulent structures, often reducing fidelity in higher-order diagnostics. More recently, Implicit Neural Networks (INRs) have been explored as continuous and grid-free field representations~\cite{sitzmann2020implicit,mildenhall2021nerf,tancik2020fourier}. Although flexible, such models typically rely on black-box parameterizations, making it difficult to interpret or explicitly control how complex flow structures are represented. Unlike these approaches, a representation that is both explicit and spatially localized can offer direct control over how flow structures are encoded.

These limitations motivate representations that are both continuous and spatially localized. An effective formulation should combine compactness with spatial adaptivity while allowing direct evaluation of derived flow quantities. This naturally suggests parametric representations built from localized basis functions, in which complex flow fields are described as a superposition of spatially concentrated primitives. Related ideas appear in wavelet decompositions, radial basis function (RBF) approximations, and mesh-free methods, all of which exploit localized kernels to represent spatial structure~\cite{mallat2002theory,fasshauer2007choosing,buhmann2003radial}. Unlike wavelet bases, which rely on fixed multiresolution decompositions, such parametric formulations allow both the location and support of basis functions to adapt directly to the flow.

In this work, we consider a Gaussian-based parametric representation of turbulent velocity fields, in which the flow is approximated as a superposition of localized kernels with learnable positions, amplitudes, and characteristic scales. Gaussian primitives have recently gained significant attention in other domains, particularly in computer graphics for efficient scene reconstruction~\cite{kerbl2023}. Related kernel-based ideas have also appeared in computational physics and fluid mechanics, for example in mesh-free and particle-based formulations where localized kernels are used to approximate field quantities and derivatives~\cite{mimeauvortex2021,purkayastha2022sph}. More recently, Gaussian-based representations have begun to appear in fluid-related contexts~\cite{gaussianXing,fluidGS2025,gaussfluid2025}. However, their use as compact, continuous parametric representations for turbulent flow fields, especially with respect to preserving small-scale and gradient-sensitive structures, remains limited and has not been systematically examined. In contrast to prior work, the present study focuses on the use of Gaussian primitives as a compact parametric representation for turbulent flow fields, with a particular emphasis on compression--fidelity trade-offs and the preservation of derivative-sensitive quantities.

Such representations offer several appealing properties for fluid dynamics. They are continuous by construction and can be evaluated at arbitrary spatial locations without requiring interpolation from a fixed grid. Their localized nature makes them well suited to spatially concentrated flow features, while their parametric form permits direct differentiation and therefore evaluation of derived quantities such as vorticity and enstrophy. In addition, explicit control over the number and distribution of kernels provides a flexible mechanism for balancing compactness and reconstruction fidelity. At the same time, it is not clear whether such localized parametric models can achieve high compression while preserving the small-scale structures that dominate turbulent dynamics.

The present study investigates this question using three-dimensional Taylor--Green vortex fields spanning different stages of flow evolution, from relatively smooth configurations to fully developed turbulence~\cite{brachet1983tgv}. We first examine the baseline isotropic Gaussian formulation and quantify how kernel count influences reconstruction accuracy and compression behavior. Reconstruction quality is assessed not only through the velocity field itself, but also through higher-order diagnostics that are sensitive to velocity gradients. These experiments reveal both the strengths and limitations of the baseline representation: while compact Gaussian models can recover the velocity field with high accuracy, their fidelity degrades in localized high-gradient regions associated with small-scale turbulent structures.

Motivated by these observations, we then investigate several structure-aware extensions designed to better capture such features. These include adaptive kernel placement guided by flow-dependent indicators, anisotropic kernels capable of aligning with local structures, and multi-resolution kernel sets that combine coarse and fine spatial scales. In addition, we examine the effect of the basis function itself by considering compact-support alternatives to Gaussian primitives. The impact of these modifications is evaluated on the most demanding flow configurations, with particular emphasis on the recovery of derivative-sensitive structures.

The main contributions of this work are as follows. First, we introduce and systematically evaluate a continuous localized Gaussian-kernel representation for compact encoding of turbulent velocity fields. Second, we assess its performance not only in terms of velocity reconstruction, but also through diagnostics sensitive to the preservation of turbulent structure. Third, we show that structure-aware extensions, particularly anisotropic kernels, improve the recovery of localized flow features and gradient-dependent quantities. Taken together, these results demonstrate that localized parametric representations provide a viable middle ground between global modal compression and neural field models, combining continuity, interpretability, and explicit control over the compression--accuracy trade-off.

\section{Methodology}

\subsection{Gaussian Parametric Representation}

We represent the fluid flow field as a continuous parametric function defined by a superposition of localized kernels. Given a spatial location $\mathbf{x} \in \mathbb{R}^3$, the velocity field $\hat{\mathbf{u}}(\mathbf{x})$ is approximated as

\begin{equation}
\hat{\mathbf{u}}(\mathbf{x}) = \sum_{i=1}^{N} w_i(\mathbf{x})\,\mathbf{a}_i,
\end{equation}
where $N$ is the number of kernels, $\mathbf{a}_i \in \mathbb{R}^3$ denotes the amplitude (velocity contribution) of the $i$-th kernel, $\boldsymbol{\mu}_i \in \mathbb{R}^3$ denotes its center (spatial location), and the kernel weights $w_i(\mathbf{x})$ are obtained from localized Gaussian responses.

In the baseline formulation used in this work, the Gaussian response is defined as

\begin{equation}
G(\mathbf{x}; \boldsymbol{\mu}_i, \boldsymbol{\sigma}_i)
=
\exp\left(
-(\mathbf{x} - \boldsymbol{\mu}_i)^T
\mathbf{D}_i^{-1}
(\mathbf{x} - \boldsymbol{\mu}_i)
\right),
\end{equation}
where $\boldsymbol{\sigma}_i = (\sigma_{ix}, \sigma_{iy}, \sigma_{iz})$ denotes axis-aligned kernel widths, and $\mathbf{D}_i = \mathrm{diag}(\sigma_{ix}^2, \sigma_{iy}^2, \sigma_{iz}^2)$. The final spatial weights are normalized across kernels according to

\begin{equation}
w_i(\mathbf{x})
=
\frac{G(\mathbf{x}; \boldsymbol{\mu}_i, \boldsymbol{\sigma}_i)}
{\sum_{j=1}^{N} G(\mathbf{x}; \boldsymbol{\mu}_j, \boldsymbol{\sigma}_j) + \varepsilon},
\end{equation}
with $\varepsilon$ a small constant for numerical stability.

This formulation provides a continuous representation of the flow field that is independent of the underlying grid and can be evaluated at arbitrary spatial locations. While similar Gaussian primitive representations have been explored in computer graphics~\cite{kerbl2023}, the present formulation differs in both purpose and usage, as it targets physically meaningful flow fields rather than visual appearance.

A key advantage of this representation is that it provides a compact parametric description of the flow field from which both primary variables and derived quantities can be computed. In particular, quantities such as vorticity and enstrophy can be obtained directly from the reconstructed field, enabling assessment of how well the compressed representation preserves derivative-sensitive flow structures.

The parameters $\{\mathbf{a}_i, \boldsymbol{\mu}_i, \boldsymbol{\sigma}_i\}_{i=1}^{N}$ are learned from data, enabling the model to adjust kernel placement and support to the underlying flow field. The number and spatial distribution of kernels provide an explicit mechanism to control the compression--accuracy trade-off. A smaller kernel budget yields higher compression but may reduce fidelity in fine-scale regions, whereas increasing the number of kernels improves reconstruction accuracy at additional computational cost.

In the present study, we restrict attention to the velocity field. Nevertheless, the same representation principle can be applied to other flow variables, such as density or pressure, by fitting the kernel amplitudes to the corresponding target quantity.

\subsection{Optimization and Training}

The parameters of the kernel representation, $\{\mathbf{a}_i, \boldsymbol{\mu}_i, \boldsymbol{\sigma}_i\}_{i=1}^{N}$, are learned by minimizing the discrepancy between the reconstructed field and reference data.

In practice, spatial coordinates are affine-normalized to the unit domain prior to optimization.

Given a set of sampled spatial locations $\{\mathbf{x}_j\}_{j=1}^{M}$ with corresponding reference velocity values $\mathbf{u}^{ref}(\mathbf{x}_j)$, we define the reconstruction loss as

\begin{equation}
\mathcal{L}_{u}
=
\frac{\sum_{j=1}^{M} \left\| \hat{\mathbf{u}}(\mathbf{x}_j) - \mathbf{u}^{ref}(\mathbf{x}_j) \right\|^2}
{\sum_{j=1}^{M} \left\| \mathbf{u}^{ref}(\mathbf{x}_j) \right\|^2 + \varepsilon}.
\end{equation}

This data fidelity term is the primary objective used throughout this work. While the framework allows the inclusion of additional loss terms based on derived quantities (e.g., vorticity or enstrophy), all results presented here use data-only training, with such quantities evaluated post-training as diagnostics.

Training is performed using randomly sampled spatial points from the reference flow field. This sampling strategy provides efficient coverage of the domain while remaining compatible with localized kernel refinement.

All kernel parameters are optimized using gradient-based methods with automatic differentiation. Kernel amplitudes are always trainable, while kernel centers and widths are also optimized to allow the representation to adapt to the underlying flow structure.

To ensure numerical stability and meaningful parameter ranges, kernel widths are constrained to remain positive and bounded, and kernel centers are restricted to the spatial domain. In practice, this is achieved through bounded parameterizations together with weak regularization applied to kernel centers and widths relative to their initialization.

This optimization framework enables the representation to adapt both the placement and scale of kernels to the underlying flow, forming the basis for the structure-aware extensions introduced in the following section.

\subsection{Structure-Aware Extensions}

While the baseline Gaussian representation provides a compact and continuous approximation of the flow field, its axis-aligned and uniformly initialized kernels may struggle to capture highly localized and anisotropic structures, particularly in turbulent regimes. To address these limitations, we introduce several structure-aware extensions that improve the representation of challenging flow features.

\paragraph{Adaptive Kernel Placement}

In regions where the reconstruction error is large, a greater fraction of the kernel budget is redirected to increase local representational capacity. In practice, an initial coarse fit is first performed using a regular kernel layout, after which an error indicator is computed over the field. A subset of kernels is then redistributed according to this error distribution, while the remaining kernels preserve global coverage. This adaptive placement allows the model to allocate resources where they are most needed, improving accuracy in localized high-gradient regions without increasing the total number of kernels.

\paragraph{Anisotropic Gaussian Kernels}

To better capture elongated flow structures such as vortical filaments and shear layers, we extend the diagonal Gaussian formulation to anisotropic kernels. Specifically, the axis-aligned width parameter is replaced by a full covariance matrix $\boldsymbol{\Sigma}_i$, leading to the kernel definition

\begin{equation}
G(\mathbf{x}; \boldsymbol{\mu}_i, \boldsymbol{\Sigma}_i)
=
\exp\left(
-(\mathbf{x} - \boldsymbol{\mu}_i)^T
\boldsymbol{\Sigma}_i^{-1}
(\mathbf{x} - \boldsymbol{\mu}_i)
\right).
\end{equation}

This formulation allows each kernel to adapt its orientation and spatial extent to align with local flow structures, providing a more efficient representation of anisotropic features.

\paragraph{Multi-resolution Kernel Representation}

To account for the wide range of spatial scales present in turbulent flows, we employ a multi-resolution strategy in which the fixed kernel budget is divided into coarse and fine subsets. Coarser kernels are intended to capture the large-scale structure of the flow, while finer kernels provide additional local resolution for smaller-scale features. In the present implementation, this is achieved by assigning different initial support widths to different subsets of kernels, optionally combined with adaptive placement.

\paragraph{Alternative Basis Functions}

In addition to Gaussian primitives, we investigate the use of alternative localized basis functions to assess the impact of kernel shape on reconstruction quality. In particular, we consider a compact-support Beta basis, in which the kernel is defined through a separable Beta-like profile with bounded support.

For a normalized local coordinate $\boldsymbol{\xi}$ defined relative to the kernel center and width, the one-dimensional Beta profile is written as

\begin{equation}
B(\xi; \alpha_i)
=
\xi^{\alpha_i - 1}(1-\xi)^{\alpha_i - 1},
\end{equation}
where a symmetric Beta profile is used and $\alpha_i > 1$ is a shape parameter controlling the concentration of the kernel within its compact support. In practice, the local coordinate is shifted and scaled so that $\xi_d \in (0,1)$ within the support of the kernel, and the profile is set to zero outside that support.

The multidimensional kernel is constructed as a separable product across spatial dimensions,

\begin{equation}
\Phi(\mathbf{x}; \boldsymbol{\mu}_i, \mathbf{s}_i, \boldsymbol{\alpha}_i)
=
\prod_{d=1}^{3}
B(\xi_d; \alpha_{i,d}),
\end{equation}

This formulation yields a compact-support kernel that is more localized than a Gaussian and is therefore of interest for representing sharp or highly concentrated structures. This increased locality, however, can introduce additional sensitivity in optimization and may lead to less smooth reconstructions if the support widths or shape parameters become overly localized.

Together, these extensions provide a flexible framework for adapting both the distribution and form of the underlying primitives, enabling improved recovery of physically significant flow structures across a range of regimes.

\section{Results}

\subsection{Baseline Evaluation}
\label{sec:baseline}

This subsection evaluates the baseline Gaussian representation on the Taylor--Green vortex, focusing on compression performance and its ability to capture turbulence-relevant quantities.

\subsubsection{Problem Setup: Taylor--Green Vortex}

We evaluate the proposed Gaussian parametric representation on the incompressible three-dimensional Taylor--Green vortex (TGV) at Reynolds number $1600$ on a uniform $256^3$ grid. This canonical benchmark captures the transition from large-scale organized flow to fully developed small-scale turbulence.

The dataset consists of velocity fields sampled at multiple time instances spanning distinct stages of flow evolution, from smooth initial conditions to states dominated by fine-scale vortical structures.

\begin{figure}
    \centering
    \includegraphics[width=0.98\textwidth]{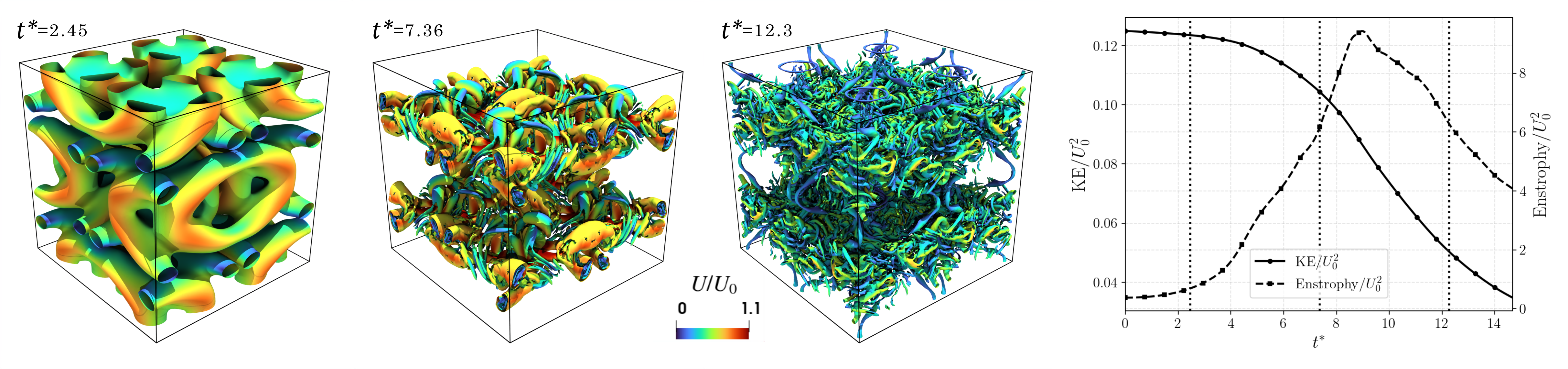}
    \caption{Evolution of the Taylor--Green vortex. Left to right: $Q$-criterion iso-surfaces at $t^*=2.45$, $7.36$, and $12.27$, representing early, transitional, and turbulent stages. Iso-surfaces are colored by normalized velocity magnitude. Right: temporal evolution of kinetic energy and enstrophy, with vertical lines indicating the selected snapshots.}
    \label{fig:tgv_overview}
\end{figure}

Figure~\ref{fig:tgv_overview} illustrates representative flow structures at selected time instances using $Q$-criterion iso-surfaces, together with the temporal evolution of kinetic energy and enstrophy. The chosen snapshots correspond to distinct stages of the flow, ranging from smooth, large-scale organization to fully developed small-scale turbulence.

For each snapshot, the Gaussian representation is trained using randomly sampled spatial points. The number of kernels $N$ is varied to study the trade-off between representation capacity and compression. Reconstruction quality is assessed using both primary variables (velocity) and derivative-based quantities (vorticity and enstrophy).

The following subsections analyze how reconstruction quality evolves across flow regimes and kernel budgets, with particular emphasis on the contrast between accurate recovery of the velocity field and the greater sensitivity of derivative-based quantities.

\subsubsection{Compression and Reconstruction Accuracy}

We first examine the relationship between compression and reconstruction accuracy for the proposed Gaussian representation. The parametric model stores only the kernel parameters $\{\mathbf{a}_i, \boldsymbol{\mu}_i, \boldsymbol{\sigma}_i\}_{i=1}^{N}$, resulting in a significantly reduced memory footprint compared to the full grid-based representation of the flow field.

To quantify this trade-off, the number of kernels $N$ is varied and reconstruction error is evaluated with respect to the reference solution. Compression is defined relative to the storage required for the original discretized velocity field.

\begin{figure}
    \centering
    \includegraphics[width=0.98\textwidth]{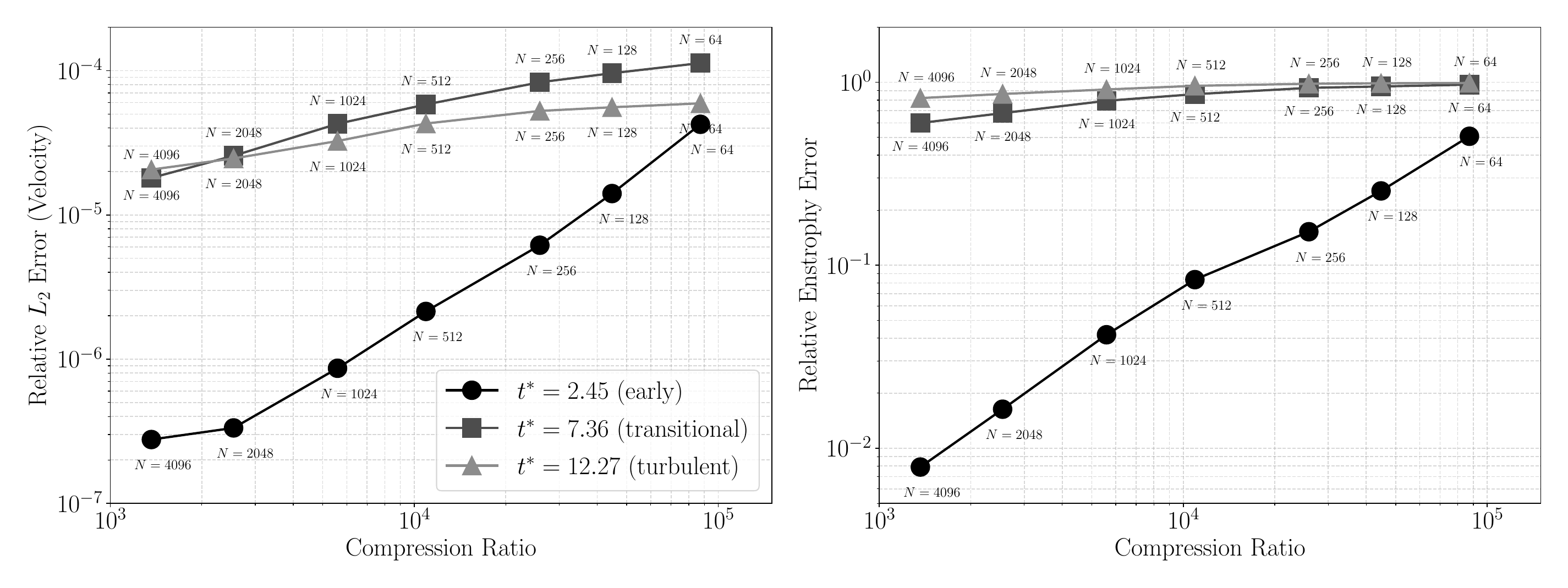}
    \caption{Compression-accuracy trade-off for the Gaussian representation at different stages of the Taylor--Green vortex. Left: relative $L_2$ error of the velocity field as a function of compression ratio. Right: relative enstrophy error. Each curve corresponds to a different time instance representing increasing flow complexity. While velocity reconstruction error decreases consistently with an increasing number of kernels, the enstrophy error remains significantly higher, particularly at later stages of the flow, highlighting the difficulty of preserving small-scale turbulent structures under strong compression.}
    \label{fig:pareto}
\end{figure}

Figure~\ref{fig:pareto} presents the resulting compression-accuracy trade-offs for three representative stages of the Taylor--Green vortex. In all cases, increasing the number of kernels improves velocity reconstruction, with errors steadily decreasing as compression is relaxed. In the early stage of the flow, characterized by smooth large-scale structures, the Gaussian representation achieves very low reconstruction error even at high compression ratios.

As the flow evolves and smaller scales emerge, reconstruction becomes increasingly challenging. While velocity errors continue to decrease with increasing kernel count, the enstrophy error remains comparatively large, particularly in later stages. In the most turbulent regime, the enstrophy error remains of order unity across a wide range of compression ratios, indicating a persistent loss of high-wavenumber content despite accurate reconstruction of the velocity field.

\begin{table}
\centering
\caption{Compression and reconstruction accuracy for the baseline Gaussian representation at representative kernel counts and flow stages. The table reports the compression ratio, relative $L_2$ velocity error, and relative enstrophy error. The original discretized velocity field contains $50{,}331{,}648$ single-precision scalar values (approximately $192$ MiB), whereas the parametric representation requires only tens to hundreds of kilobytes.}
\label{tab:baseline_tradeoff}
\begin{tabular}{ccccccc}
\toprule
$t^{*}$ & $N$ & Compression Ratio & \# Parameters & Compressed Size (KiB) & $L_2$ Error & Rel. Enstrophy Error \\
\midrule
2.45  & 512  & $1.09\times10^{4}$ & 4608  & 18.0  & $2.13\times10^{-6}$ & 0.120 \\
2.45  & 2048 & $2.55\times10^{3}$ & 19773 & 77.2  & $3.33\times10^{-7}$ & 0.041 \\
2.45  & 4096 & $1.37\times10^{3}$ & 36864 & 144.0 & $2.86\times10^{-7}$ & 0.028 \\
\midrule
7.36  & 512  & $1.09\times10^{4}$ & 4608  & 18.0  & $5.81\times10^{-5}$ & 0.878 \\
7.36  & 2048 & $2.55\times10^{3}$ & 19773 & 77.2  & $2.61\times10^{-5}$ & 0.687 \\
7.36  & 4096 & $1.37\times10^{3}$ & 36864 & 144.0 & $1.79\times10^{-5}$ & 0.601 \\
\midrule
12.27 & 512  & $1.09\times10^{4}$ & 4608  & 18.0  & $4.31\times10^{-5}$ & 0.961 \\
12.27 & 2048 & $2.55\times10^{3}$ & 19773 & 77.2  & $2.70\times10^{-5}$ & 0.882 \\
12.27 & 4096 & $1.37\times10^{3}$ & 36864 & 144.0 & $2.20\times10^{-5}$ & 0.836 \\
\bottomrule
\end{tabular}
\end{table}

Table~\ref{tab:baseline_tradeoff} provides representative numerical values. While velocity errors can be reduced to very low levels even at compression ratios exceeding $10^{4}$, the corresponding enstrophy errors remain of order unity in later turbulent stages, indicating a persistent loss of small-scale information.

These results highlight a key limitation of the baseline isotropic Gaussian formulation: although it is highly effective at compressing the velocity field, it struggles to faithfully represent derivative-sensitive quantities associated with localized turbulent structures. This limitation is not solely due to insufficient model capacity; it is intrinsic to the isotropic Gaussian representation, which lacks the ability to align with anisotropic, filamentary flow structures.

\subsubsection{Temporal Evolution of Energy and Enstrophy}

\begin{figure}
    \centering
    \includegraphics[width=0.98\textwidth]{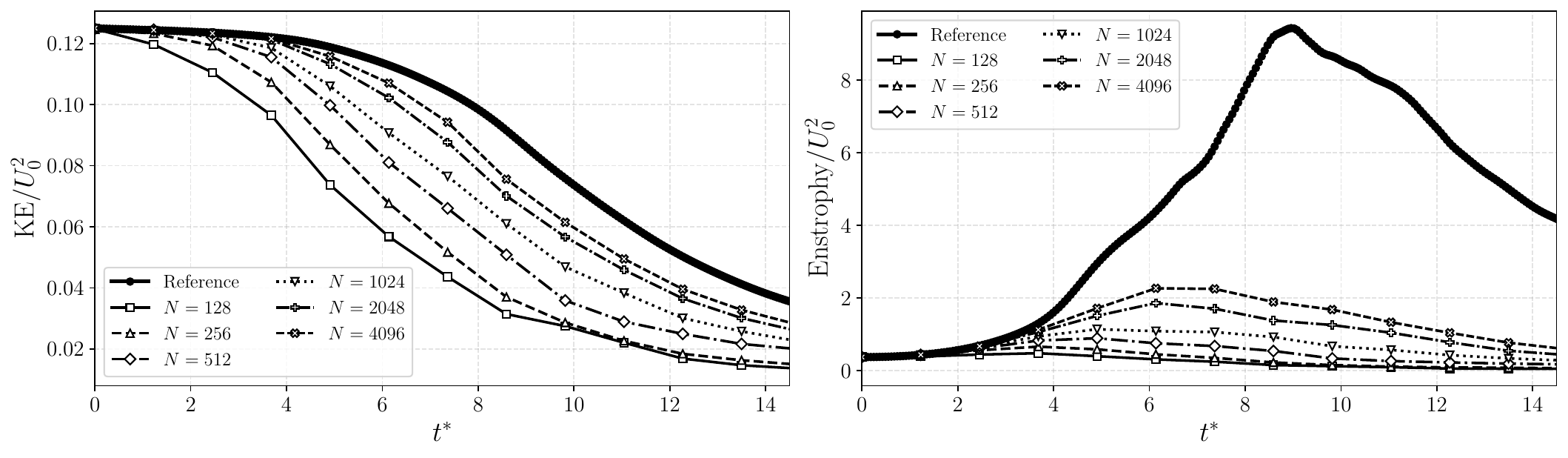}
    \caption{Temporal evolution of kinetic energy (left) and enstrophy (right) for the reference solution and Gaussian reconstructions with varying numbers of kernels. While kinetic energy is captured with reasonable accuracy and converges with increasing $N$, enstrophy is significantly underpredicted across all cases, particularly during the peak turbulent phase. This indicates that the representation preserves large-scale dynamics more effectively than derivative-sensitive small-scale turbulent structures.}
    \label{fig:ke_enstrophy}
\end{figure}

Figure~\ref{fig:ke_enstrophy} shows the temporal evolution of kinetic energy and enstrophy for the reference solution and the reconstructed fields.

Kinetic energy is well captured across all kernel counts, with increasing $N$ improving agreement with the reference solution. The overall decay trend and magnitude are preserved, indicating that the dominant large-scale dynamics are retained.

In contrast, enstrophy is consistently underpredicted. Even at the highest resolution ($N=4096$), the peak enstrophy and its subsequent decay are not accurately reproduced. The discrepancy is most pronounced during the peak turbulent phase, where enstrophy is dominated by fine-scale vortical structures.

This behavior is consistent with the spatial observations in Figure~\ref{fig:recon_slices}. Kinetic energy is dominated by large-scale motions and is therefore relatively insensitive to small-scale smoothing, whereas enstrophy depends on velocity gradients and is highly sensitive to the loss of high-wavenumber content. In this sense, the Gaussian representation effectively acts as a low-pass filter, attenuating the small-scale features responsible for enstrophy production.

Increasing the number of kernels improves the reconstruction of large-scale quantities, but does not fully recover the missing small-scale dynamics. This indicates that the limitation is intrinsic to the form of the representation, rather than solely due to insufficient kernel count.

\subsubsection{Spatial Localization of Reconstruction Error}

\begin{figure}
    \centering
    \includegraphics[width=0.98\textwidth]{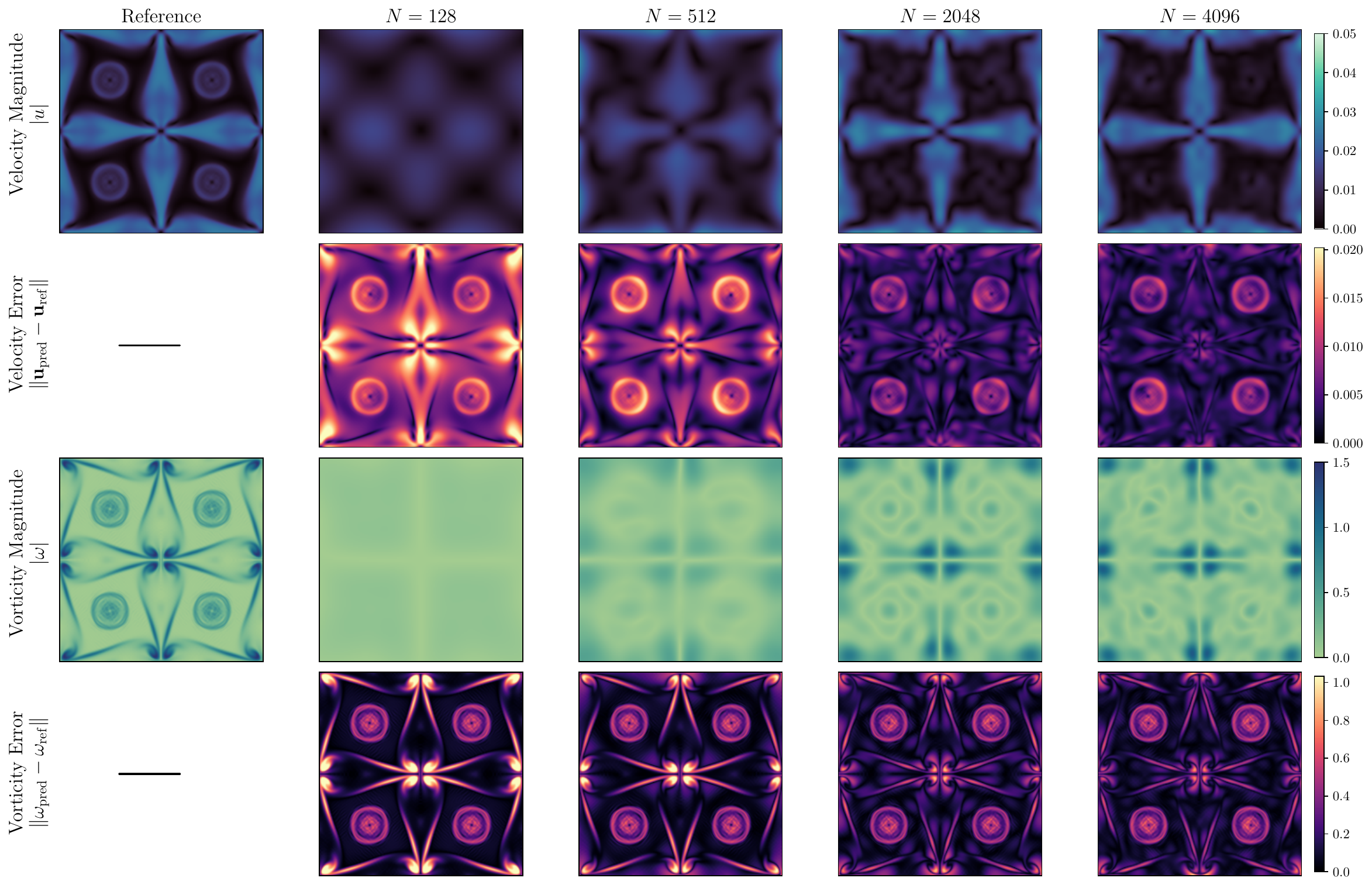}
    \caption{Comparison of reference and reconstructed fields at $t^* = 12.27$ for increasing Gaussian kernel budgets. Shown are representative two-dimensional slices of the velocity magnitude (first row), velocity absolute error (second row), vorticity magnitude (third row), and vorticity absolute error (fourth row). The first column shows the reference solution, while the remaining columns correspond to reconstructions with increasing numbers of kernels. As the kernel count increases, the large-scale velocity field is recovered accurately, whereas fine-scale vortical structures remain substantially more difficult to reconstruct.}
    \label{fig:recon_slices}
\end{figure}

Figure~\ref{fig:recon_slices} compares representative two-dimensional slices of the velocity magnitude and vorticity magnitude across increasing kernel budgets at the most challenging turbulent snapshot, $t^*=12.27$. The reference field is shown in the first column, followed by reconstructions obtained with increasing numbers of Gaussian kernels. The corresponding absolute errors are shown in the second and fourth rows.

The velocity field is reconstructed with good fidelity even at relatively low kernel counts and improves progressively as the number of kernels increases. The dominant large-scale structures and overall flow organization are preserved across all cases, consistent with the low relative $L_2$ errors reported earlier.

The vorticity field exhibits a markedly different behavior. Although increasing the kernel budget improves the reconstruction, fine-scale vortical structures remain substantially less accurate than the velocity field. Thin filaments and regions of sharp gradients are smeared, weakened, or entirely absent at low and moderate kernel counts, and remain imperfect even at the largest budget considered.

The corresponding error maps show that the reconstruction error is not distributed uniformly throughout the domain. Instead, it is concentrated in localized regions associated with strong gradients and coherent vortical structures. This indicates that the baseline Gaussian representation is effective at recovering the large-scale organization of the flow, but systematically under-resolves localized high-gradient features. In particular, isotropic kernels are unable to align with the anisotropic geometry of filamentary structures, leading to concentrated reconstruction errors in these regions.

This spatial error localization is consistent with the spectral attenuation observed earlier, linking the loss of high-wavenumber content to specific flow structures.


\subsubsection{Interpretation: Scale Separation in Gaussian Representations}

The preceding results reveal a consistent scale-dependent behavior in the Gaussian representation.

The Gaussian representation efficiently captures low-wavenumber components of the flow, corresponding to large-scale coherent structures, but attenuates high-wavenumber content associated with small-scale turbulence.

As a result, quantities dominated by large scales, such as kinetic energy, are accurately preserved, while gradient-sensitive quantities such as enstrophy are significantly degraded.

This behavior is intrinsic to isotropic Gaussian kernels, which cannot align with anisotropic, filamentary flow structures. The representation therefore introduces a systematic bias toward smooth solutions, leading to a loss of turbulence-relevant small-scale features.

In this sense, the Gaussian representation behaves as an implicit low-pass filter, fundamentally limiting its ability to represent fully developed turbulent flows.

\subsection{Comparison with Standard Representations}
\label{sec:standard_representation_comparison}

To contextualize the Gaussian representation against more conventional alternatives, we compare the most challenging Taylor--Green vortex snapshot ($t^*=12.27$) with two additional single-snapshot baselines: a sparse wavelet representation and a neural implicit representation based on SIREN. All methods are evaluated on the same velocity field and compared using the same diagnostics, namely the relative velocity reconstruction error, the relative enstrophy error, and the effective compression ratio.

While the preceding analysis isolates the limitations of the Gaussian representation, it remains important to assess how these limitations compare to alternative compact representations under similar storage constraints.

The baseline Gaussian method corresponds to the diagonal Gaussian representation with normalized kernel blending and $N=4096$ kernels. The wavelet baseline uses a three-dimensional Daubechies-4 (\texttt{db4}) wavelet transform~\cite{daubechies1992} applied component-wise, retaining the largest $24{,}576$ coefficients. This corresponds to $49{,}152$ stored values when both coefficient magnitudes and indices are counted. Coefficients are selected by magnitude, retaining the largest entries across all components, and reconstruction is performed using the inverse transform. The SIREN baseline~\cite{sitzmann2020implicit} is trained directly on spatial coordinates and velocity values for the same snapshot, and the reported model uses $50{,}435$ trainable parameters. Additional training details for the SIREN baseline are provided in Appendix~\ref{app:training_details}. These baselines are chosen so that all three methods operate in a comparable storage regime, ensuring that differences in performance reflect representational behavior rather than model size.

\begin{figure}
    \centering
    \includegraphics[width=0.98\textwidth]{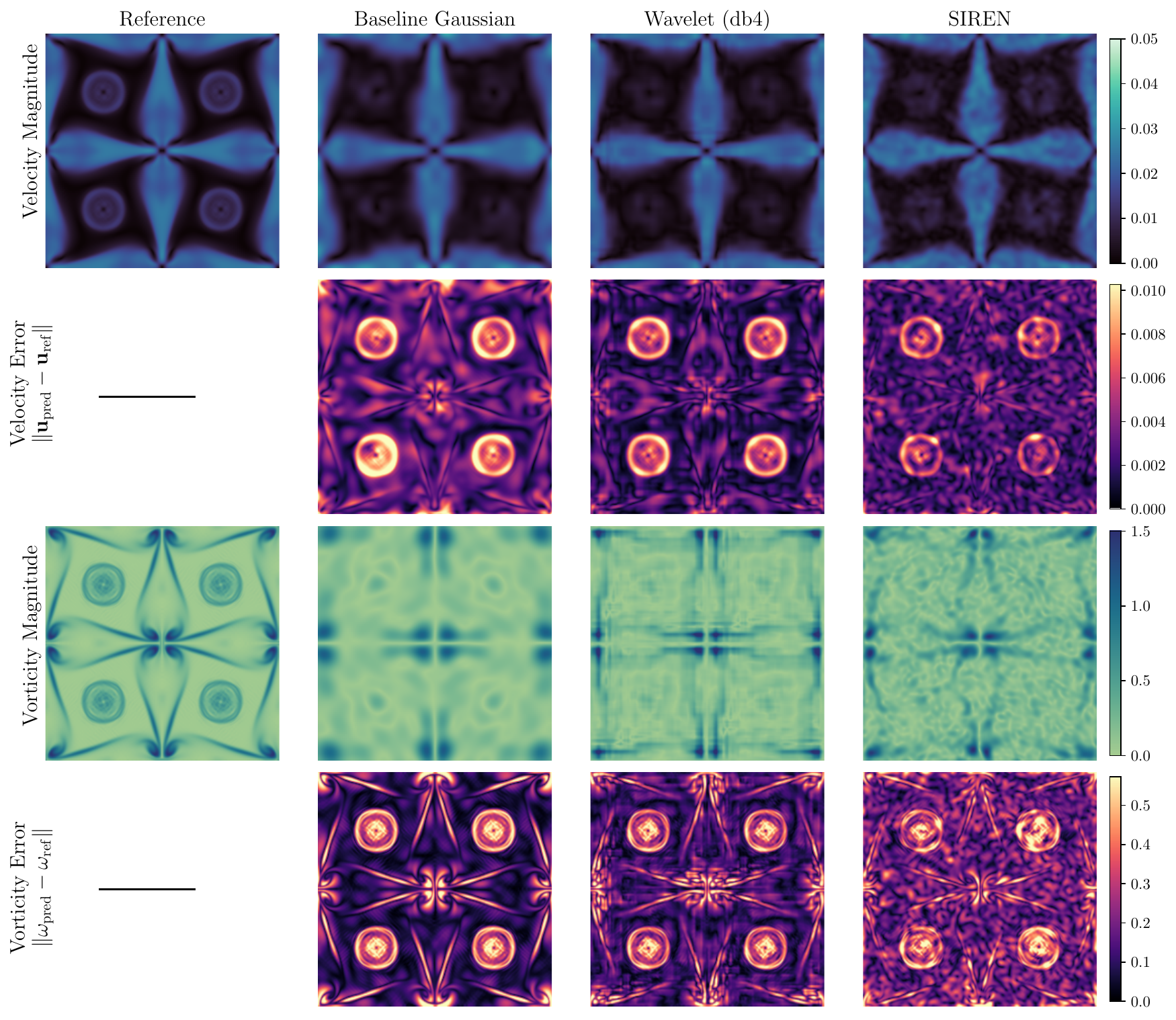}
    \caption{Single-snapshot comparison at $t^*=12.27$ between the reference field, the Gaussian representation, a \texttt{db4} wavelet baseline, and a SIREN baseline under comparable storage constraints. Rows show velocity magnitude, velocity error, vorticity magnitude, and vorticity error. The Gaussian representation provides a smooth and accurate reconstruction of the velocity field, but under-resolves fine-scale vortical structures. The wavelet baseline preserves sharper gradients and more coherent structures, while the SIREN baseline exhibits larger reconstruction errors and less coherent spatial organization. All error maps use a shared color scale to enable consistent comparison across methods.}
    \label{fig:standard_representation_comparison}
\end{figure}

Figure~\ref{fig:standard_representation_comparison} provides a qualitative comparison of the reconstructed fields. The Gaussian representation produces the smoothest reconstruction of the velocity field, while the wavelet baseline achieves slightly lower reconstruction error and preserves more fine-scale vortical structures. The SIREN baseline performs least favorably, with visibly larger reconstruction errors.

These observations are quantified in Table~\ref{tab:single_snapshot_baselines}. The Gaussian method achieves a very low relative $L_2$ error of $2.06\times10^{-5}$ at a compression ratio of $1365.3\times$, demonstrating strong performance in compact value-space reconstruction. However, the corresponding enstrophy error remains high ($8.20\times10^{-1}$), suggesting the loss of small-scale turbulent structures observed in previous sections.

\begin{table}[t]
    \centering
    \caption{Comparison of the most challenging Taylor--Green vortex snapshot ($t^*=12.27$) using the Gaussian method, a \texttt{db4} wavelet baseline, and a SIREN baseline. Reported are the relative velocity reconstruction error, the relative enstrophy error, the number of stored values or parameters, and the resulting compression ratio.}
    \label{tab:single_snapshot_baselines}
    \begin{tabular}{lcccc}
        \hline
        Method & Relative $L_2$ Error & Relative Enstrophy Error & Stored Values & Compression Ratio \\
        \hline
        Baseline Gaussian & $2.06\times10^{-5}$ & $8.20\times10^{-1}$ & $36{,}864$ & $1365.3\times$ \\
        Wavelet (\texttt{db4}) & $1.07\times10^{-5}$ & $5.79\times10^{-1}$ & $49{,}152$ & $1024.0\times$ \\
        SIREN & $1.30\times10^{-1}$ & $6.47\times10^{-1}$ & $50{,}435$ & $998.0\times$ \\
        \hline
    \end{tabular}
\end{table}

In contrast, the \texttt{db4} wavelet baseline achieves both a lower velocity reconstruction error ($1.07\times10^{-5}$) and a significantly improved enstrophy error ($5.79\times10^{-1}$), indicating stronger retention of high-wavenumber content. This improvement comes at a slightly reduced compression ratio ($1024.0\times$) and with less smooth spatial reconstruction, as seen in Figure~\ref{fig:standard_representation_comparison}.

The SIREN baseline performs substantially worse in this setting, with a relative $L_2$ error of $1.30\times10^{-1}$ and no competitive advantage in enstrophy or compression. Despite providing a continuous representation, it does not yield a favorable trade-off compared to either the Gaussian or wavelet approaches.

\begin{figure}
    \centering
    \includegraphics[width=0.98\textwidth]{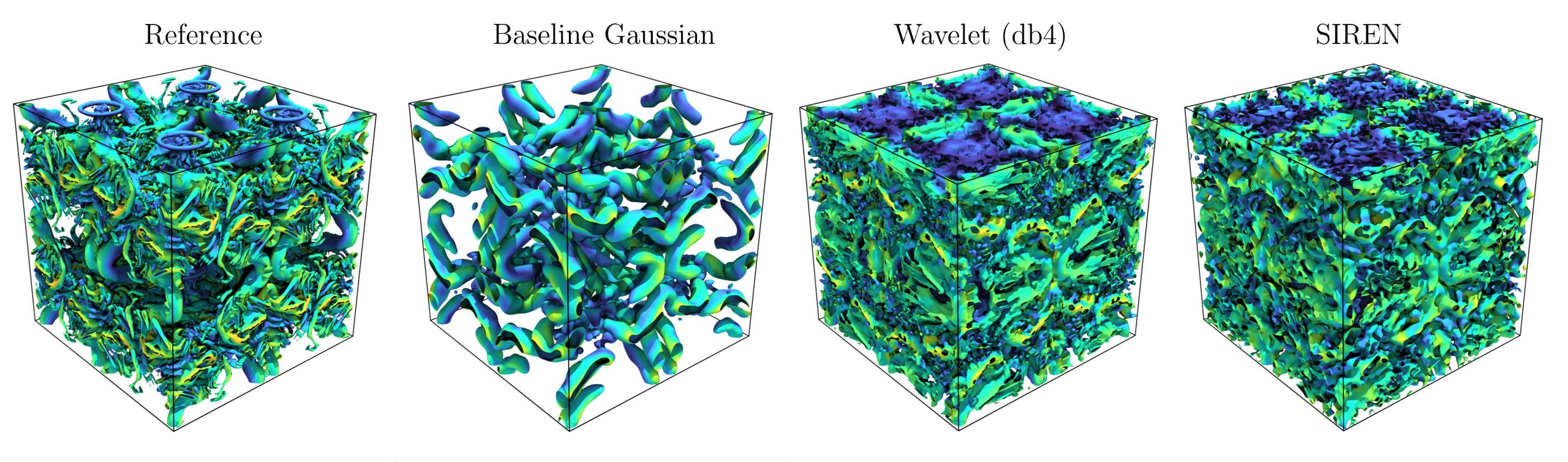}
    \caption{Comparison of coherent vortical structures at $t^* = 12.27$ using $Q$-criterion isosurfaces for the reference solution, the Gaussian representation, a \texttt{db4} wavelet baseline, and a SIREN baseline. All methods are evaluated under comparable storage constraints and use the same $Q$ threshold. Isosurfaces are colored by normalized velocity magnitude $|\mathbf{u}|/U_0$. The Gaussian representation produces significantly fewer and smoother structures, indicating a loss of small-scale vortical topology. The wavelet baseline preserves a richer set of coherent structures but introduces mild spatial artifacts, while the SIREN baseline exhibits noisier and less coherent structures.}
    \label{fig:standard_representation_comparison_Q}
\end{figure}

While these metrics quantify reconstruction accuracy, they do not fully capture differences in physically meaningful flow structures. To further assess the ability of each representation to preserve coherent vortical structures, we examine $Q$-criterion isosurfaces in Figure~\ref{fig:standard_representation_comparison_Q}.

The $Q$-criterion visualization highlights a key difference between representations that is not fully captured by global error metrics. While the Gaussian model achieves low velocity reconstruction error, it significantly under-resolves coherent vortical structures, producing fewer and smoother isosurfaces compared to the reference. In contrast, the wavelet baseline retains a denser and more connected vortex network, consistent with its improved enstrophy accuracy, albeit with some visible artifacts. The SIREN reconstruction exhibits less coherent and noisier structures, reflecting its overall lower accuracy.

Overall, the comparison reveals a clear trade-off between smooth reconstruction and preservation of derivative-sensitive structures. The Gaussian representation excels at compact and accurate reconstruction of the velocity field, whereas the wavelet baseline better preserves small-scale features relevant to enstrophy and flow topology.

Importantly, while wavelets provide stronger fixed-basis representations of small-scale content, the Gaussian formulation offers a spatially adaptive parametric representation with explicit control over kernel location, scale, and geometry, enabling targeted extensions such as anisotropy and adaptive placement that are not naturally accessible in fixed-basis methods. This distinction is particularly relevant for turbulent flows, where structures are highly localized and anisotropic.

These results suggest that improving Gaussian-based representations requires increasing their structural expressiveness rather than simply increasing parameter count. The wavelet and SIREN baselines considered here should therefore be viewed as parameter-matched reference methods rather than exhaustively optimized competitors. Their role is to provide context for the strengths and limitations of the Gaussian formulation under a comparable storage budget, and to motivate the structure-aware extensions introduced in the following section.

\subsection{Structure-Aware Kernel Extensions}

The preceding results reveal a fundamental limitation of the baseline Gaussian representation: it under-resolves derivative-sensitive quantities despite accurately reconstructing the velocity field. This limitation is intrinsic to isotropic kernels, which cannot align with elongated vortical structures and thin shear layers that dominate small-scale turbulence.

To address this limitation, we introduce a set of structure-aware extensions designed to increase the expressiveness of the Gaussian representation while preserving its compact, continuous, and parametric nature. These extensions target complementary aspects of the representation, including spatial allocation, kernel anisotropy, and multi-scale structure. The goal is not to increase the number of parameters, but to use the existing kernel budget more effectively by better aligning the representation with the underlying flow physics.

\subsubsection{Comparison Through Temporal Energy and Enstrophy Evolution}

\begin{figure}
    \centering
    \includegraphics[width=0.98\textwidth]{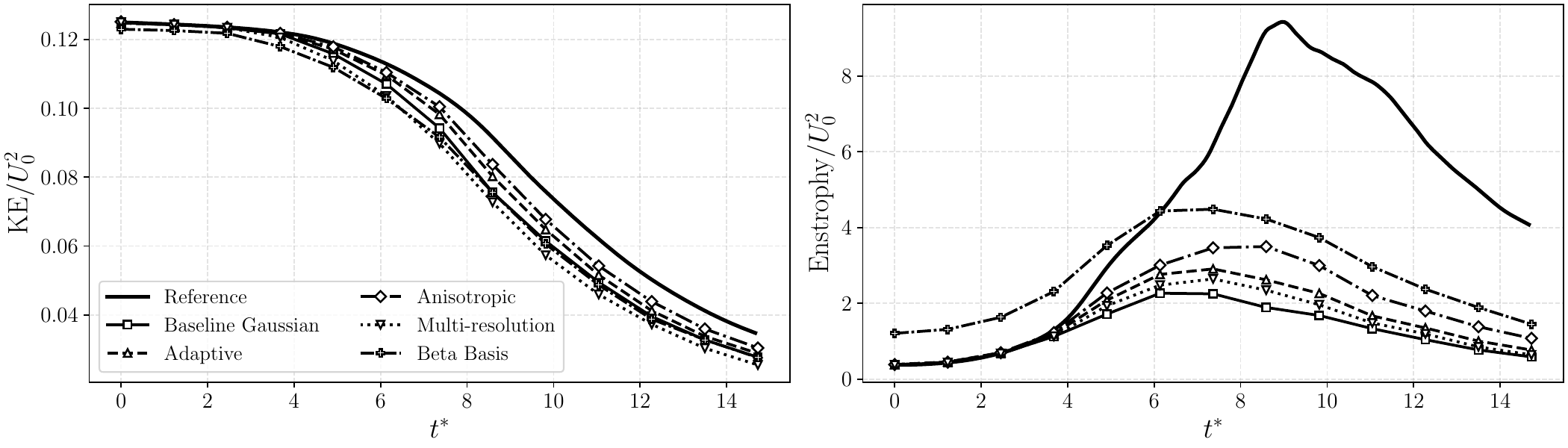}
    \caption{Temporal evolution of kinetic energy (left) and enstrophy (right) for the reference solution, the baseline Gaussian representation, and the tested structure-aware kernel extensions at fixed kernel budget ($N=4096$). All methods reproduce the kinetic-energy decay with reasonable accuracy, indicating that the large-scale flow dynamics are preserved. Larger differences appear in the enstrophy evolution, which is more sensitive to fine-scale structures. Adaptive placement and multi-resolution kernels provide only modest improvement over the baseline, whereas the anisotropic Gaussian formulation yields the best overall balance between energy preservation and recovery of derivative-sensitive flow features. The Beta basis improves enstrophy relative to the baseline in some cases, but remains less accurate overall and introduces visible spatial artifacts.}
    \label{fig:method_comparison}
\end{figure}

To compare the proposed extensions under identical conditions, all variants are evaluated at a fixed kernel budget of $N=4096$. Figure~\ref{fig:method_comparison} shows the temporal evolution of kinetic energy and enstrophy for the reference solution, the baseline Gaussian representation, and the tested structure-aware variants.

All methods reproduce kinetic energy well, indicating that large-scale dynamics are preserved.

In contrast, the enstrophy evolution exhibits much larger discrepancies. The baseline Gaussian model significantly underpredicts enstrophy, especially during the peak turbulent phase, consistent with the loss of fine-scale vortical structures observed in the spatial reconstructions.

The adaptive placement strategy yields only limited improvement over the baseline, indicating that redistributing kernels alone is insufficient to recover the missing small-scale dynamics. The multi-resolution formulation similarly provides only modest gains, suggesting that a simple coarse--fine decomposition does not fully address the loss of derivative-sensitive structures under the present configuration.

Among the tested extensions, the anisotropic Gaussian formulation provides a clear and consistent improvement over the baseline. By allowing each kernel to adopt a full covariance structure, the representation can stretch and align with the local geometry of the flow, enabling more effective coverage of elongated vortical structures and thin shear layers that are poorly captured by isotropic kernels. This directly improves recovery of derivative-sensitive quantities without increasing parameter count. Spectral analysis further suggests improved retention of intermediate and high-wavenumbers.

The Beta basis increases enstrophy in some cases but introduces visible artifacts and degrades overall reconstruction quality; it is therefore treated as exploratory rather than competitive.

Overall, these results demonstrate that increasing kernel expressiveness through anisotropy is more effective than modifying spatial allocation alone.

\begin{table}[t]
    \centering
    \caption{Relative enstrophy error at the most challenging Taylor--Green vortex snapshot ($t^*=12.27$) for the baseline and the tested structure-aware extensions. Errors are reported in percent.}
    \label{tab:structure_aware_enstrophy}
    \begin{tabular}{lcc}
        \hline
        Method & Relative Enstrophy Error [\%] & Compression Ratio \\
        \hline
        Baseline Gaussian & 82.00 & $1365.3\times$ \\
        Adaptive Gaussian & 78.51 & $1365.3\times$ \\
        Multi-resolution Gaussian & 81.01 & $1365.3\times$ \\
        Anisotropic Gaussian & 71.42 & $1024.0\times$ \\
        Beta Basis & 62.24 & $1024.0\times$ \\
        \hline
    \end{tabular}
\end{table}

\subsubsection{Spectral Comparison}

\begin{figure}
    \centering
    \includegraphics[width=0.98\textwidth]{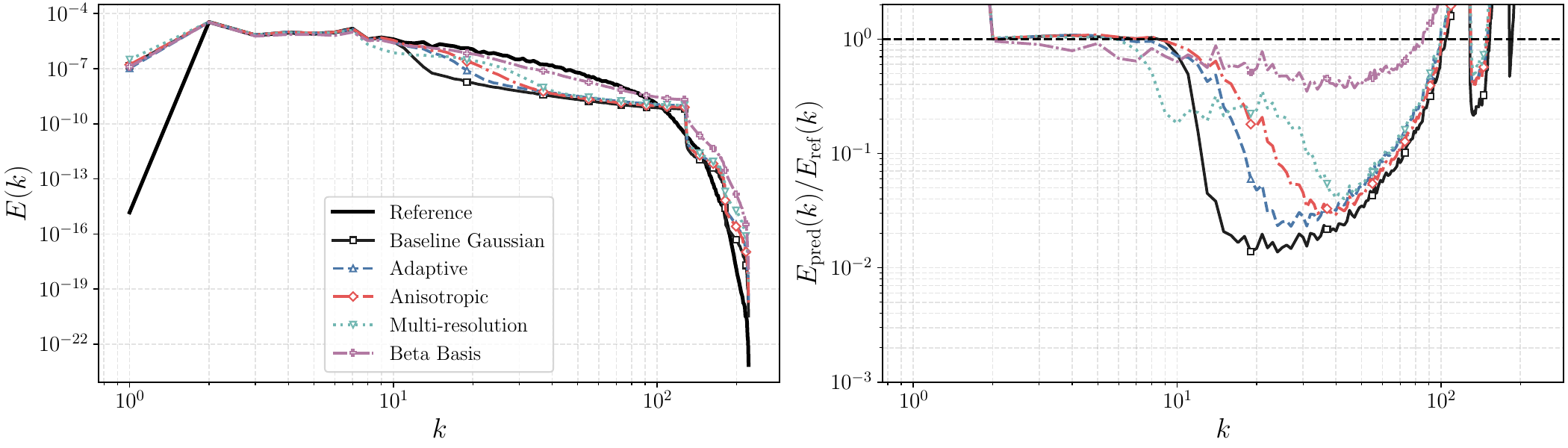}
    \caption{Spectral comparison at $t^*=12.27$ for the reference solution, the baseline Gaussian representation, and structure-aware kernel extensions at fixed kernel budget ($N=4096$). The left panel shows the kinetic-energy spectrum $E(k)$, while the right panel shows the spectral retention ratio $E_{\mathrm{pred}}(k)/E_{\mathrm{ref}}(k)$. While most methods reproduce the low-wavenumber behavior, the ratio plot reveals strong attenuation of intermediate- and high-wavenumber content in the baseline representation. Among the tested variants, the anisotropic Gaussian formulation retains the largest fraction of spectral energy across these ranges.}
    \label{fig:method_spectrum}
\end{figure}

Figure~\ref{fig:method_spectrum} compares the kinetic-energy spectrum at $t^*=12.27$ for the baseline Gaussian representation and the tested structure-aware extensions. The left panel shows the absolute spectrum, while the right panel reports the retained fraction of spectral energy relative to the reference field. This second representation makes the loss of small-scale energy more explicit than the absolute spectrum alone.

The baseline model reproduces the low-wavenumber behavior reasonably well, but its retained spectral energy decreases rapidly with increasing wavenumber, indicating substantial smoothing of intermediate- and high-wavenumber content. This behavior is consistent with the known sensitivity of high-wavenumber content to smoothing operations and reinforces the interpretation of the Gaussian representation as an implicit low-pass filter. The adaptive and multi-resolution variants modify this trend only modestly, consistent with their limited improvement in enstrophy and spatial reconstruction.

By contrast, the anisotropic formulation preserves a significantly larger fraction of the reference spectrum across the intermediate- and high-wavenumber ranges. This behavior is consistent with its improved recovery of elongated vortical structures and its lower enstrophy error. The Beta basis also changes the spectral distribution, but this spectral change is accompanied by visible reconstruction artifacts, highlighting that improved spectral retention alone does not guarantee accurate spatial reconstruction.

Overall, the spectral analysis reinforces that improving kernel expressiveness through anisotropy is more effective than redistributing kernels or combining multiple support scales. These results indicate that the primary limitation of the baseline Gaussian representation is not the number of kernels, but the geometric rigidity of isotropic kernels in representing anisotropic turbulent structures.

\section{Conclusions}

In this work, we introduced a continuous Gaussian parametric representation for compact encoding of three-dimensional turbulent flow fields. The proposed approach achieves substantial compression while maintaining high velocity reconstruction accuracy, demonstrating the potential of localized kernel-based representations as practical alternatives to grid-based storage.

Through systematic evaluation on the Taylor--Green vortex, we showed that the Gaussian representation exhibits clear scale-dependent behavior. Large-scale flow structures and kinetic energy are accurately preserved, whereas derivative-sensitive quantities such as vorticity and enstrophy are significantly degraded. This behavior arises from the intrinsic low-pass filtering effect of isotropic Gaussian kernels, which are unable to align with anisotropic, filamentary turbulent structures.

Comparisons with wavelet and neural implicit baselines further highlighted this limitation. While the wavelet baseline better preserves small-scale features under comparable storage constraints, the Gaussian formulation provides a spatially adaptive parametric representation with explicit control over kernel location and geometry, enabling targeted extensions beyond fixed-basis methods.

To address the observed limitations, we introduced a set of structure-aware extensions, including adaptive kernel placement, multi-resolution kernels, and anisotropic Gaussian kernels. Among these, the anisotropic formulation provided the most consistent improvement in recovering derivative-sensitive quantities, demonstrating that increasing kernel expressiveness is more effective than merely increasing parameter count or redistributing kernels alone.

Taken together, these results indicate that the primary limitation of Gaussian-based representations for turbulent flows lies in geometric expressiveness rather than capacity. Future work will focus on combining anisotropic kernels with adaptive placement strategies and physics-informed criteria to further improve the representation of small-scale turbulent structures, moving toward a fully structure-aware and physically grounded continuous representation of turbulent flows.


\section*{CRediT authorship contribution statement}

Dhanush V. Shenoy: Conceptualization, Methodology, Software, Validation, Investigation, Formal analysis, Visualization, Writing -- original draft, Writing -- review \& editing.

Steven H. Frankel: Supervision, Writing -- review \& editing.

\section*{Declaration of Competing Interest}

The authors declare that they have no known competing financial interests or personal relationships that could have appeared to influence the work reported in this paper.

\section*{Code Availability}

The implementation used in this study is publicly available at \href{https://github.com/Dhanushenoy/gaussian-flow-codecs}{https://github.com/Dhanushenoy/gaussian-flow-codecs}.

\section*{Acknowledgments}

The authors acknowledge the CFDLAB at Technion -- Israel Institute of Technology for providing computational resources and support.

\bibliographystyle{plainnat}
\bibliography{ref}

\appendix

\section{Training Details}
\label{app:training_details}

This appendix summarizes the training configuration used throughout the manuscript. Unless otherwise stated, all models share the same optimization procedure and differ only in kernel parameterization or initialization strategy.

\subsection{Data and Preprocessing}

All flow fields are read from structured-grid VTK files. Spatial coordinates $\mathbf{x}\in\mathbb{R}^3$ are affinely normalized independently along each axis to the unit domain,
\[
\tilde{\mathbf{x}} = \frac{\mathbf{x}-\mathbf{x}_{\min}}{\mathbf{x}_{\max}-\mathbf{x}_{\min}},
\]
while velocity values are kept in their original units.

Training is performed on randomly sampled spatial points from the grid, and all evaluation metrics are computed on the full grid after reconstruction.

\subsection{Kernel Representation}

The velocity field is approximated as
\[
\hat{\mathbf{u}}(\mathbf{x}) = \sum_{i=1}^{N} w_i(\mathbf{x})\,\mathbf{a}_i,
\]
where $\mathbf{a}_i\in\mathbb{R}^3$ are kernel amplitudes and $w_i(\mathbf{x})$ are normalized spatial weights,
\[
w_i(\mathbf{x}) = \frac{\phi_i(\mathbf{x})}{\sum_{j=1}^{N}\phi_j(\mathbf{x})+\varepsilon},
\]
with $\varepsilon=10^{-12}$.

Each kernel is parameterized by a center $\boldsymbol{\mu}_i\in[0,1]^3$, a scale parameter, and an amplitude vector. Baseline models use diagonal Gaussian kernels, while the structure-aware extensions modify the covariance structure, spatial distribution, or basis function.

\subsection{Optimization}

All models are trained using the Adam optimizer with a base learning rate of $10^{-2}$ and minibatches of $10{,}000$ randomly sampled spatial points. The primary loss is the relative $L_2$ reconstruction error,
\[
\mathcal{L}_{\mathrm{data}} =
\frac{\|\hat{\mathbf{u}} - \mathbf{u}\|_2^2}{\|\mathbf{u}\|_2^2 + \varepsilon}.
\]

Unless otherwise stated, no physics-aware loss is used during training; vorticity and enstrophy are evaluated only as post-training diagnostics.

Kernel amplitudes are always trainable. Kernel centers and widths are also optimized, using learning-rate multipliers of $10^{-2}$ and $5\times10^{-2}$, respectively, unless otherwise stated. Weak regularization is applied to kernel centers and widths relative to their initialization, with regularization weights of $10^{-6}$ for both. The model parameters corresponding to the lowest training loss are retained.

All experiments were repeated with multiple random initializations, and no significant variation in the reported trends was observed.

\subsection{Baseline Configuration}

Baseline experiments use Gaussian kernels with diagonal covariance, normalized blending, and regular-grid initialization. Kernel centers and widths are trainable. Kernel counts span
\[
N \in \{64,\,128,\,256,\,512,\,1024,\,2048,\,4096\},
\]
and each model is trained for $4000$ optimization steps.

For the baseline Gaussian experiments, kernel widths are constrained to
\[
10^{-3} \le \sigma \le 0.5.
\]

\subsection{Structure-Aware Extensions}

All structure-aware variants are evaluated at a fixed kernel budget of $N=4096$ using the same data-only training objective and $2000$ optimization steps. Adaptive placement redistributes part of the kernel budget toward regions of high reconstruction error, guided by a short coarse probe. In the present implementation, the adaptive fraction is $0.5$, with a global width scale of $1.0$, a local width scale of $0.5$, and a coarse probe of $50$ optimization steps. The anisotropic variant replaces diagonal Gaussian kernels with full covariance parameterizations, allowing kernels to align with local flow structures.

The multi-resolution variant introduces coarse and fine kernel subsets using a fine-kernel fraction of $0.5$, together with coarse and fine width scales of $1.5$ and $0.5$, respectively. The Beta variant replaces Gaussian kernels with compact-support Beta basis functions. In the final reported Beta runs, the Beta shape parameter is fixed to $3.0$ rather than optimized in order to improve training stability. The Beta width learning-rate multiplier is reduced to $2\times10^{-2}$, and the minimum kernel width is increased to $5\times10^{-3}$.

\subsection{Parameter Constraints}

To ensure numerical stability, kernel centers are constrained to the unit domain using a sigmoid parameterization, and kernel widths are restricted to a bounded interval using a softplus-based parameterization and clipping. Full covariance matrices are parameterized through a bounded Cholesky factorization.

\subsection{Compression Accounting}

Compression ratios are computed relative to the number of scalar values in the original velocity field. For the Gaussian models, the stored representation consists of kernel centers, kernel shape parameters, and kernel amplitudes. For diagonal Gaussian kernels, this corresponds to three center coordinates, three width parameters, and three amplitude values per kernel. For anisotropic kernels, the diagonal width parameters are replaced by a full covariance parameterization. For Beta kernels, the additional shape parameters are included in the stored count.

For the wavelet baseline, storage includes both retained coefficient values and their indices after truncation of the three-dimensional \texttt{pywt} Daubechies-4 transform. For the SIREN baseline, storage is defined as the total number of trainable network parameters. Reported compression ratios therefore correspond to parameter-count and coefficient-count comparisons, rather than serialized file sizes on disk.

\subsection{Evaluation Metrics}

After training, reconstructed fields are evaluated on the full grid. Reported metrics include:
\begin{itemize}
\item relative $L_2$ velocity error,
\item vorticity diagnostics,
\item relative enstrophy error,
\item compression ratio.
\end{itemize}

\subsection{Summary of Key Hyperparameters}

Table~\ref{tab:appendix_hyperparameters} summarizes the principal hyperparameters used in the baseline and structure-aware experiments.

\begin{table}[t]
    \centering
    \caption{Summary of key training hyperparameters used in the reported experiments.}
    \label{tab:appendix_hyperparameters}
    \begin{tabular}{lcc}
        \hline
        Setting & Baseline Gaussian & Structure-Aware Variants \\
        \hline
        Kernel budget & $N \in \{64,\dots,4096\}$ & $N=4096$ \\
        Training steps & $4000$ & $2000$ \\
        Batch size & $10{,}000$ & $10{,}000$ \\
        Optimizer & Adam & Adam \\
        Base learning rate & $10^{-2}$ & $10^{-2}$ \\
        Center LR multiplier & $10^{-2}$ & $10^{-2}$ \\
        Width LR multiplier & $5\times10^{-2}$ & $5\times10^{-2}$ \\
        Physics loss & none & none \\
        Blend mode & normalized & normalized \\
        \hline
    \end{tabular}
\end{table}

\section{Illustration of Gaussian Kernel Primitives}
\label{app:gaussian_primitives}

\begin{figure}
    \centering
    \includegraphics[width=\textwidth]{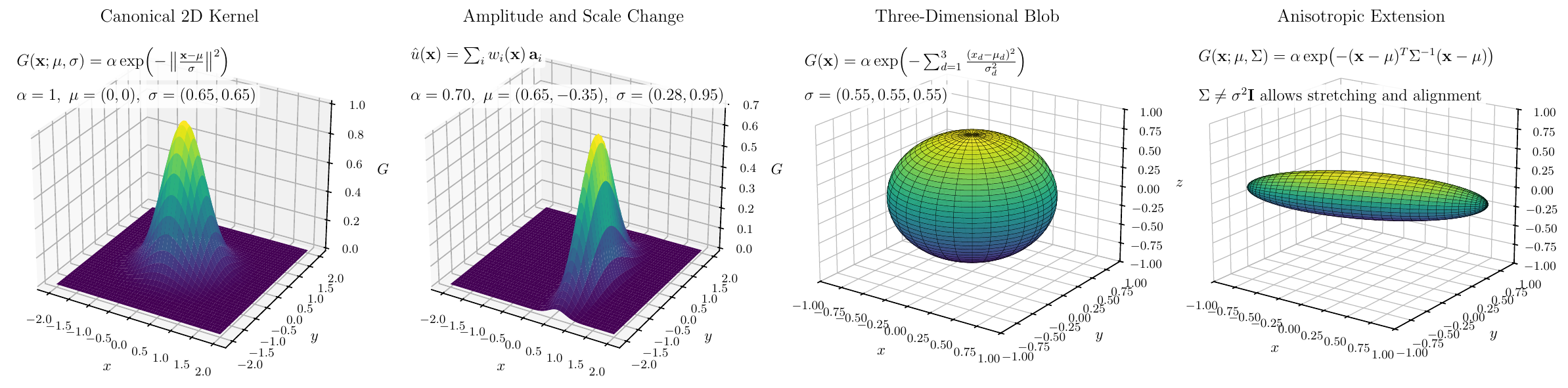}
    \caption{Illustration of the Gaussian kernel primitives used in the proposed representation. The first column shows a canonical two-dimensional Gaussian kernel. The second column illustrates the effect of changing the kernel amplitude, center, and support, producing a shifted and anisotropically scaled kernel. The third column shows the corresponding three-dimensional isotropic kernel as a compact blob. The fourth column illustrates the anisotropic extension, in which the diagonal support parameter is replaced by a full covariance matrix, allowing the kernel to stretch and align with local flow structures.}
    \label{fig:gaussian_primitives}
\end{figure}

Figure~\ref{fig:gaussian_primitives} illustrates the Gaussian primitives underlying the proposed representation. The examples show how changes in amplitude, center, and support modify the kernel in two dimensions, and how the extension from isotropic to anisotropic kernels alters the corresponding three-dimensional structure.

\section{Additional Training Diagnostics}
\label{app:training_diagnostics}

\begin{figure}
    \centering
    \includegraphics[width=\textwidth]{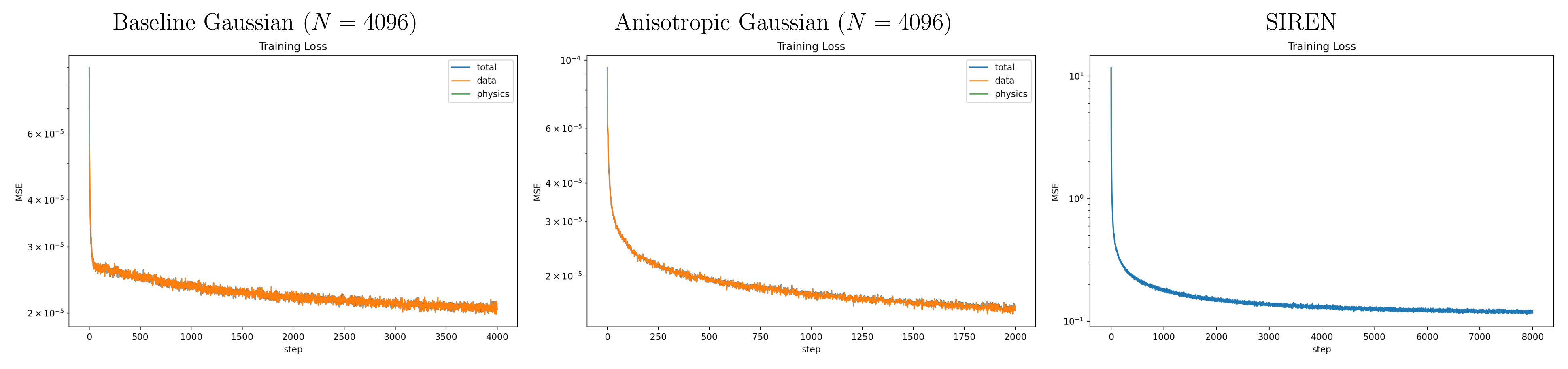}
    \caption{Representative training curves for the baseline Gaussian model, the anisotropic Gaussian extension, and the SIREN baseline for the most challenging Taylor--Green vortex snapshot ($t^*=12.27$). The Gaussian-based models converge stably to low loss values, whereas the SIREN baseline remains at substantially higher loss throughout training. The wavelet baseline is not shown because it is constructed directly using \texttt{pywt} transform truncation rather than iterative optimization.}
    \label{fig:training_curve_comparison}
\end{figure}

Figure~\ref{fig:training_curve_comparison} compares representative training curves for the baseline Gaussian model, the anisotropic Gaussian extension, and the SIREN baseline on the most challenging Taylor--Green vortex snapshot ($t^*=12.27$). These curves are included to document the optimization behavior of the learned parametric representations under the training settings used in this work.

The Gaussian-based models exhibit stable optimization and converge rapidly to low reconstruction loss. The anisotropic formulation achieves a lower final loss than the baseline Gaussian model, consistent with its improved reconstruction quality reported in the main text. In contrast, the SIREN baseline converges much more slowly and remains at a substantially higher loss level, reflecting the weaker reconstruction performance observed in the single-snapshot comparison.

The wavelet baseline is not included in this figure because it is not obtained through iterative optimization. In the present work, the wavelet representation is constructed directly using the \texttt{pywt} implementation of the three-dimensional Daubechies-4 transform, followed by coefficient truncation. It therefore does not admit a training curve comparable to those of the Gaussian and SIREN models.

\section{Sequence-Level Comparison with POD}
\label{sec:pod_comparison}

To further contextualize the proposed representation, we compare it with a sequence-level reduced-order baseline based on proper orthogonal decomposition (POD)~\cite{berkooz1993proper}. Unlike the Gaussian, wavelet, and SIREN comparisons presented earlier, which operate on individual snapshots, POD is constructed from the full Taylor--Green vortex time series and therefore exploits correlations across time through a shared linear basis.

In the present implementation, the velocity snapshots are assembled into a matrix and decomposed using singular value decomposition. Truncated POD models are then formed using a small number of retained modes, and each snapshot is reconstructed from the corresponding modal coefficients. To provide a more informative storage estimate, the cost of the shared POD basis and mean field is amortized across the full sequence, and the reported per-snapshot storage additionally includes the modal coefficients for that snapshot.

\begin{figure}
    \centering
    \includegraphics[width=0.9\textwidth]{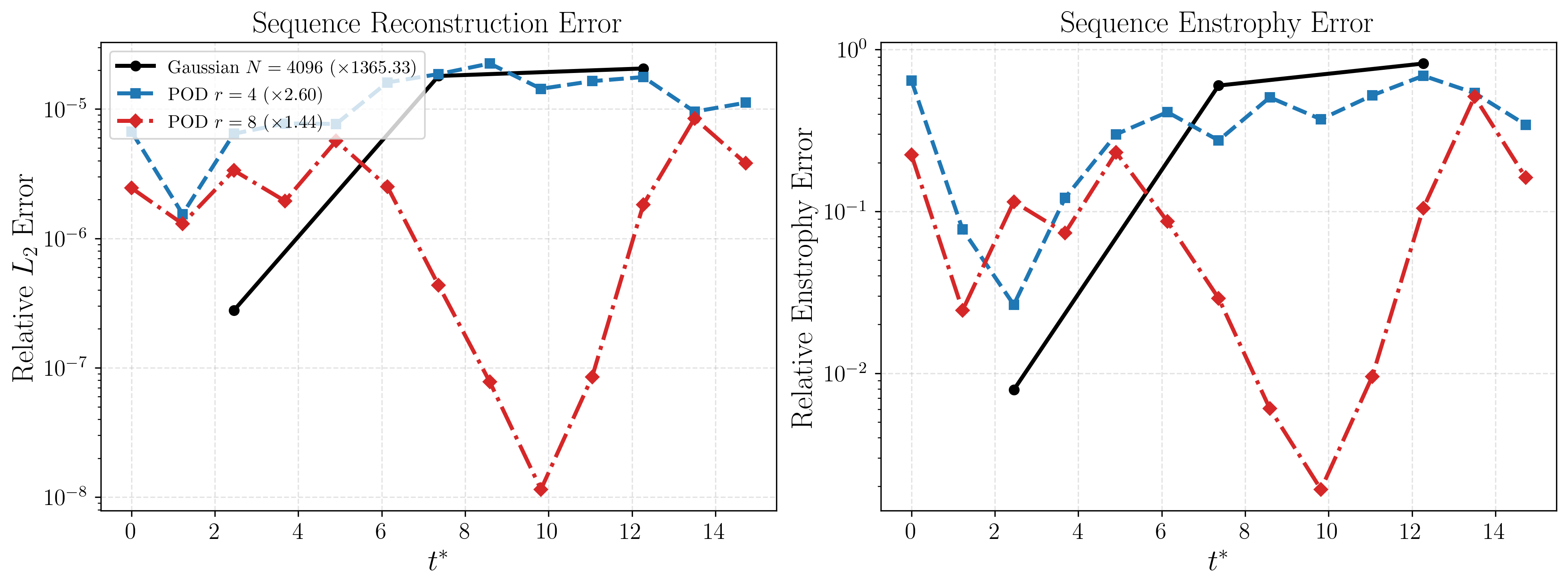}
    \caption{Sequence-level comparison between the baseline Gaussian model ($N=4096$) and POD models with ranks $r=4$ and $r=8$ over the Taylor--Green vortex time series. The left panel shows the relative velocity reconstruction error, and the right panel shows the relative enstrophy error. POD provides substantially stronger reconstruction quality on this short correlated sequence, but with much lower effective per-snapshot compression than the Gaussian representation.}
    \label{fig:pod_sequence_comparison}
\end{figure}

Figure~\ref{fig:pod_sequence_comparison} compares the temporal evolution of the relative velocity reconstruction error and the relative enstrophy error for the baseline Gaussian model with $N=4096$ and POD models with ranks $r=4$ and $r=8$. As expected, POD provides very strong reconstruction quality on this short sequence. In particular, the rank-$8$ model achieves substantially smaller velocity and enstrophy errors than the baseline Gaussian model over most of the time interval. Even the rank-$4$ model remains competitive with the Gaussian baseline in velocity error and improves the reconstruction of enstrophy during the turbulent phase.

At the same time, the compression ratios reveal an important distinction between the two approaches. Although POD is highly effective as a sequence-level reduced-order model, its amortized per-snapshot compression remains modest for the present dataset, with compression ratios of approximately $2.6\times$ for $r=4$ and $1.44\times$ for $r=8$. In contrast, the Gaussian representation provides a much more compact per-snapshot model, with a compression ratio of approximately $1365\times$ for $N=4096$, albeit with substantially larger enstrophy error.

The comparison therefore highlights a fundamental trade-off. POD serves as a strong linear reduced-order baseline when a shared basis can be learned across an entire sequence, whereas the Gaussian formulation targets compact snapshot-wise parametric representation with explicit spatial primitives. The present results show that POD can outperform the baseline Gaussian model in reconstruction quality on a short correlated sequence, but at the cost of much weaker effective compression in the per-snapshot setting considered here.

\end{document}